\documentclass[prl,twocolumn,amsmath]{revtex4}
\usepackage{graphicx}
\usepackage{dcolumn}
\usepackage{bm}

\begin{document}

\title{Hybrid optical pumping of optically dense alkali-metal vapor without
quenching gas}
\author{M. V. Romalis}
\affiliation{Department of Physics, Princeton University, Princeton,
New Jersey 08544}

\begin{abstract}
Optical pumping of an optically thick atomic vapor typically
requires a quenching buffer gas, such as N$_{2}$, to prevent
radiation trapping of unpolarized photons which would depolarize the
atoms. We show that optical pumping of a trace contamination of Rb
present in K metal results in a 4.5 times higher polarization of K
than direct optical pumping of K in the absence of N$_{2}$. Such
spin-exchange polarization transfer from optically-thin species is
useful in a variety of areas, including spin-polarized nuclear
scattering targets and electron beams, quantum-non-demolition spin
measurements, and ultra-sensitive magnetometry.
\end{abstract}

\pacs{32.80.Xx, 32.50.+d}

\maketitle

%Use showkeys class option if keyword
%display desired

% Force line breaks with \\

Optical pumping is an extremely versatile technique that can by used
to achieve nearly complete atomic polarization \cite{Hap72}.
However, in an optically-thick medium radiation trapping imposes a
limit on the maximum achievable polarization because spontaneously
emitted photons do not have the same polarization and wavelength as
the optical pumping light. Multiple reabsorption of these photons
causes depolarization and for a high optical density completely
prevents optical pumping \cite{Ander2}. It was realized soon after
development of optical pumping that molecular buffer gases can be
used to quench excited atomic states and prevent emission of
spontaneous radiation. As a result, many applications of optical
pumping, particularly for alkali-metal atoms, use N$_{2}$ buffer gas
to prevent radiation
trapping \cite{WalkerRev}. However, there are several situations where the presence of 5-50 torr of N$%
_{2}$ gas is detrimental. One example is polarization of H and D
nuclear spins by spin exchange with alkali metals for use as nuclear
targets in electron scattering experiments \cite{GaoRev,Gao}. A
large magnetic field is used in this case to reduce the effect of
radiation trapping \cite{Ander1,Anderswalker} even though it also
reduces the efficiency of spin-exchange, particularly for D
\cite{Holt}. Another example where the quenching gas is detrimental
is for polarization of an electron beam using optically-pumped
alkali-metal atoms \cite{Gay1,Gay2}.

Spin-exchange collisions can be used to transfer spin polarization
between different species, as was first demonstrated by Dehmelt
\cite{Dehmelt}. Recently ``hybrid'' optical pumping using Rb and K
atoms has been explored
for spin-exchange optical pumping of $^{3}$%
He gas \cite{WalkerSE,WalkerSE1}. Optical pumping of Rb atoms with
readily available lasers creates spin polarization in K, which is
then transferred to $^{3}$He with a higher overall efficiency than
by direct Rb-$^{3}$He spin exchange \cite{Baranga}. Here we point
out that hybrid optical pumping can be used in a wider set of
circumstances when the optical density of one of the alkali-metal
atoms is kept small. This allows optical pumping in the absence of
N$_2$ gas. In addition to the examples mentioned above, we point out
that absence of N$_{2}$ improves the effective optical depth in
quantum non-demolition spin measurements using Faraday rotation
\cite{Takahashi}.

Small optical density is also beneficial for spin-exchange
relaxation free (SERF) magnetometers \cite{SERF}. In this case the
limiting factor is not the presence of N$_{2}$ gas, but the
attenuation of the pumping beam. In steady-state operation the
magnetometer has optimal sensitivity when the optical pumping rate
is equal to the spin relaxation rate, a condition that cannot be
maintained throughout an optically thick vapor. It is possible to
use a far-detuned pump laser to reduce absorption, but this causes
unwanted light shifts \cite{Shah}. Optical pumping of a low-density
alkali metal and Faraday rotation measurements on the high-density
metal will achieve optimal sensitivity throughout the sensor.

We experimentally demonstrate spin polarization of optically dense K
vapor without buffer gas by spin exchange with optically pumped Rb.
The spin polarization of K obtained in this way is a factor of 4.5
higher than by direct optical pumping of K. Interestingly, the Rb
metal was not intentionally introduced into the cell, but was
present in the vapor at approximately 0.2\% due to contamination of
K metal. We also develop a density matrix model for optical pumping
in the presence of high spin-exchange rate but without any buffer
gas and point out particular sensitivity to the circular
polarization of the pumping light in this situation.

Consider two alkal-metal species undergoing spin-exchange collisions, one
much more abundant than the other. For definiteness we consider K and $^{85}$%
Rb, with a small Rb fraction $f=n_{Rb}/(n_{Rb}+n_{K})\ll 1$. Rb
atoms are being optically pumped with an average absorption rate $R$
for an unpolarized atom and both atoms are undergoing electron spin
randomization collisions at a rate $\Gamma$. The thermally-averaged
spin-exchange rate constant $\left\langle \sigma _{ex}v\right\rangle
$ for different alkali-metal atoms as well as their mutual
spin-exchange rates are nearly the same \cite{Gibbs,Stark}. Thus,
the spin exchange rate of atom $i$ with atom $j$ is given by
$X_{ij}=\left\langle
\sigma _{ex}v\right\rangle n_{j}$ and the total spin-exchange rate $%
X=\left\langle \sigma _{ex}v\right\rangle (n_{Rb}+n_{K}).$ If we
make a simplifying assumption of fast electron spin randomization in
the excited state, for example due to collisions with helium buffer
gas, it is particularly easy to derive a simple equation for the
spin polarization of K atoms due to optical pumping of Rb
\cite{Appelt}. For $f\ll 1$ we get
\begin{equation}
P_{K}=1/[1+\Gamma (1/R+1/X)/f].  \label{Pol}
\end{equation}
Thus, two conditions need to be satisfied to achieve a high K polarization, $%
fR\gg \Gamma $ and $fX\gg \Gamma $. The first condition ensures that
the overall input rate of angular momentum from the pump beam is
sufficient to polarize all atoms and the second condition ensures
that the spin exchange rate of K with Rb exceeds its own spin
relaxation rate.

A more detailed analysis is necessary in the absence of buffer gas,
when there is no electron randomization in the excited state and the
depopulation optical pumping is determined by spontaneous emission.
For this case we calculate the equilibrium density matrix in the
presence of spin exchange, ground state electron spin relaxation,
and optical pumping in vacuum. We assume the laser is tuned to the peak
of $F=2\rightarrow F^{\prime }\rightarrow 3$ D1 transition in $^{85}$%
Rb. Fig.~1 shows a contour plot of $P_{K}$ as a function of $fR/\Gamma$ and $%
fX/\Gamma.$ The plot is made for $f=0.01$, but the results are
virtually identical for $f=0.1-0.001$. One can see that
Eq.~(\ref{Pol}), plotted with dashed lines, gives a good
approximation to the full density matrix calculation.

\begin{figure}
\centering
\includegraphics[width=8cm]{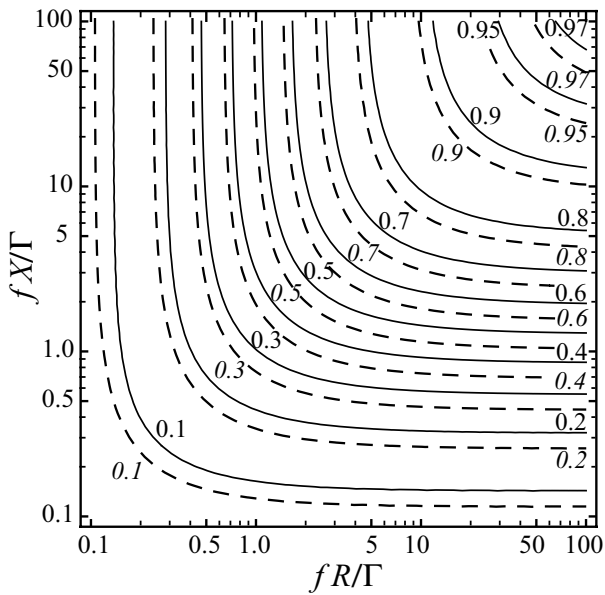}
\caption{Contour plot of maximum polarization of K as a function of
 the Rb pumping rate $R$, the spin exchange rate $X$, and the Rb fraction $f$.
 The solid lines and
 upright numbers correspond to the density matrix calculations while dashed lines and italic numbers correspond to Eq.~(\ref{Pol})
  } \label{Maxpol}
\end{figure}

In the absence of a buffer gas one typically uses an anti-relaxation
surface coating to reduce spin relaxation of alkali-metal atoms on
cell walls \cite{Robertson}. The spin relaxation rate is then given
by $ \Gamma \simeq \bar{v}/lN_e$, where $\bar{v}$ is thermal
velocity of atoms, $l$ is the characteristic  dimension of the cell,
determined by the volume/surface ratio, and $N_e$ is the number of
bounces that the coating allows on average before electron spin
relaxation. In order to avoid radiation trapping in optical pumping
of Rb we need $\sigma _{0}n_{Rb}l<1$. The K optical density
$OD_{K}=\sigma _{0}n_{K}l,$ where $\sigma _{0}$ is the peak
absorption cross-section, including the effects of Doppler
broadening, which is similar for all alkali atoms. Thus, to avoid
radiation trapping we need $f\simeq 1/OD_{K}$. It follows that
$fX/\Gamma =N_e\sigma _{ex}/\sigma _{0}$. The peak Doppler broadened
absorption cross-section in alkali-metal atoms is about
$\sigma_{0}=7\times 10^{-12}$ cm$^{2}$, while the spin exchange cross-section is $%
\sigma _{ex}=2\times 10^{-14}$ cm$^{2}$. Hence we need $N_e\gg 300$
in order to achieve high spin polarization. Paraffin has long been
used as an anti-relaxation coating with operating temperature up to
50$^{\circ}$C. It typically allows about 10000 surface bounces
before relaxation of the atomic polarization. The relaxation is
predominantly due to electron spin randomization
 \cite{Budker}, so taking into account the nuclear slowing
down factor it corresponds to $N_e\sim1700$ for an alkali atom with
$I=3/2$. More recently OTS coating with $N_e> 300$ has been shown to
operate at temperatures up to 170$^{\circ}$C \cite{Seltzer}. Alkane
coatings recently reported in \cite{Balabas1,Balabas2} allow over
$N_e\gg 10^5$ bounces and can operate up to 100$^{\circ}$C. A higher
temperature results in a higher density of the more abundant alkali
metal, which is advantageous for most applications.

We experimentally explore spin-exchange optical pumping in a 1.9
inch diameter evacuated cell with K metal and an OTS surface
coating. Previously it was found that in the absence of N$_{2}$
quenching the maximum K polarization achieved with optical pumping
of K was limited to 2-3\% \cite{Seltzer}. It has been known
anecdotally that commercial alkali-metal samples are often
cross-contaminated, so a small impurity of Rb could be present in
the cells. We have found that Rb vapor density was about $2\times
10^{-3}$ of the K density and also observed trace amounts of Cs in
the K cell. An absorption spectrum of Rb and K at 160$^{\circ}$C is
shown in Fig.~2. It can be seen that while K vapor is optically
thick, the optical density of Rb is significantly less than 1, ideal
for optical pumping in the absence of quenching gas.

\begin{figure}
\centering
\includegraphics[height=8cm,angle=90]{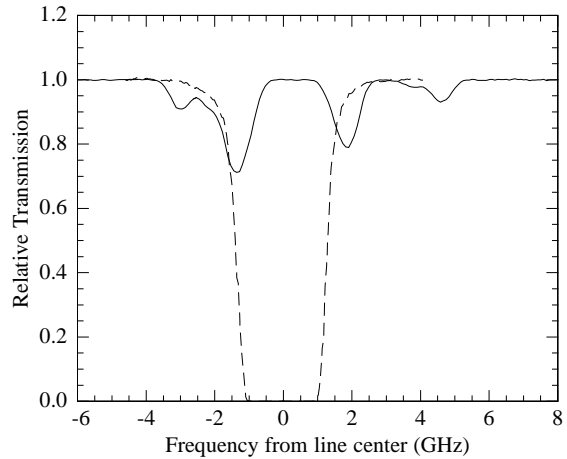}
\caption{Absorption spectrum of K (dashed line) and trace amounts of
Rb (solid line) at 160$^{\circ}$C in a K cell.}
\end{figure}

Fig.~3 shows measurements of the K spin polarization with
paramagnetic optical rotation of an off-resonant linearly polarized
laser detuned by 51 GHz from the K D1 line. The K vapor is optically
pumped by a co-propagating circularly polarized laser tuned to the
Rb or to the K D1 line and parallel to a holding magnetic field. The
maximum polarization obtained when pumping on Rb is about 4.5 times
larger than when pumping directly on K. The inset shows the power
dependance of the maximum rotation signal as a function of K pump
laser power, demonstrating the saturation and eventual decrease of
the polarization due to radiation trapping. In contrast, for Rb
pumping the K polarization continues to increase with available
laser power. For the data in Fig.~3 the K density is determined from
spin-exchange relaxation rate to be $n_{K}=9.2\times
10^{12}$~cm$^{-3}$. The optical rotation corresponds to K
polarization obtained with Rb pumping of 13.7\% vs. 3.0\% for direct
pumping on K. The time constant for the transient decay of spin
polarization in the dark is $T_1=62$ msec, which corresponds to
$N_e=120$, taking into account nuclear slowing down factor for K. We
find that both Eq.~(1) and the density matrix model predict the K
polarization with Rb pumping to be $10-15$\%, in agreement with
experimental measurements within uncertainties of the input
parameters.

\begin{figure}
\centering
\includegraphics[width=7.5cm]{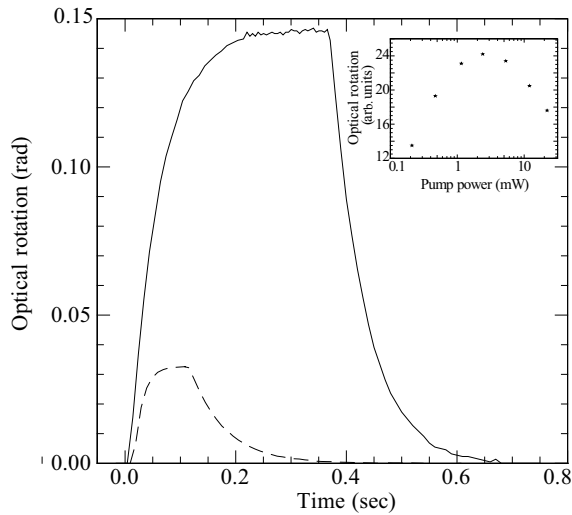}
\caption{Optical rotation transient in K vapor generated by optical
pumping of optically dense K (dashed line) or trace amounts of Rb
(solid line). The pump light is abruptly turned on and off. The
inset shows the maximum polarization of K as a function of K pumping
laser power.}
\end{figure}

The dependence of the K spin-polarization on the Rb pump laser
frequency is not trivial, as shown in Fig.~4. At a low optical
pumping rate, the spectrum of K spin polarization, shown with solid
triangles, resembles the Rb absorption spectrum. However, if the
optical pumping rate is larger than the spin-exchange rate, the
polarization exhibits a more complicated profile that is very
sensitive to the degree of light polarization. As can be seen in
Fig.~4, the measured polarization spectrum is very different from
predictions for perfectly circularly polarized light, but follows
closely the model assuming light polarization of 80\%.  The
polarization sensitivity is due to different degree of hyperfine
optical pumping. For perfectly circularly polarized light
$F=3$,$M=3$ state in $^{85}$Rb is a  dark state and Rb atoms can
reach 100\% polarization at high optical pumping intensity. However,
if the light polarization is not perfectly circular, hyperfine
pumping into $F=2$ state takes place when the optical pumping rate
exceeds the spin exchange rate. This results in the reversal of the
electron spin polarization of Rb as well as K atoms near $F=3$
$^{85}$Rb line.

\begin{figure}
\centering
\includegraphics[height=7cm]{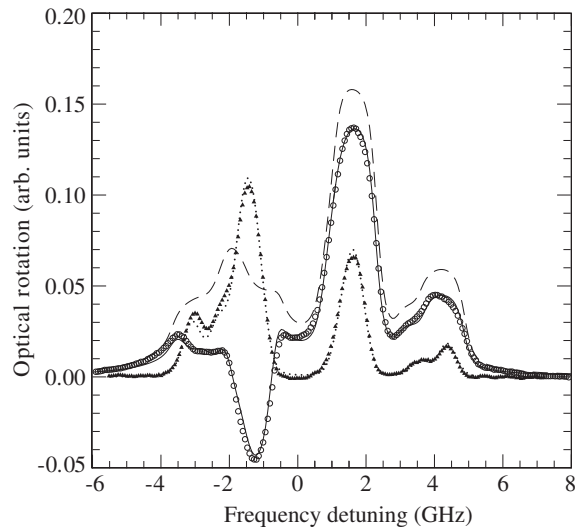}
\caption{Optical rotation in K vapor as a function of Rb pump light
frequency. Open circles show K polarization measurements for Rb
optical pumping rate much larger than the spin-exchange rate.
Triangles show the K polarization for Rb pumping rate much smaller
than the spin-exchange rate, with data scaled up by a factor of 14.
Note reversal of the polarization sign for pump light near $F=3$
$^{85}$Rb lines due to hyperfine optical pumping. The theoretical
prediction for low pumping rate is shown with dots, for high pumping
rate and perfect circular polarization with a dashed line and for
80\% circular polarization with a solid line.}
\end{figure}

Below we give  two examples beyond those already explored in the
literature where hybrid pumping with low optical density of one
species is particularly useful. In quantum-non-demolition (QND)
measurements using paramagnetic Faraday rotation \cite{Takahashi}
the intensity of the probe beam is usually limited by its absorption
rate. In the regime of far detuning of the probe laser, the wing of
the Voight absorption profile is dominated by the Lorentzian width
$\Gamma _{L}$, which is the sum of the natural atomic decay rate and
pressure broadening linewidth. The optical rotation angle is given
by $\phi =nlr_{e}cf_{osc}(\nu -\nu _{0})/[(\nu -\nu _{0})^{2}+\Gamma
_{L}^{2}]/2,$ while the absorption cross-section is $\sigma
=r_{e}cf_{osc}\Gamma _{L}/[(\nu -\nu _{0})^{2}+\Gamma _{L}^{2}]$. If
the noise in  measurements of Faraday rotation  is dominated by
photon shot noise, and spin relaxation of the atoms is limited by
the probe beam photon absorption rate,  one can show that the
signal-to-noise ratio in one
atomic spin relaxation time is equal to $\phi /\delta \phi =\sqrt{N_{at}OD_L}$%
, where $OD_{L}=r_{e}cf_{osc}nl/\Gamma _{L}.$ Hence, the atom spin
can be readout with a SNR that exceeds atom shot noise
$\sqrt{N_{a}}$ by a factor of $\sqrt{OD_{L}}$, thus allowing QND
measurements that follow atomic spin evolution. The $OD_{L}$ that
enters here corresponds to the Lorentzian linewidth, not to the
actual observed optical depth on resonance, which is dominated by
Doppler effect for hot atoms. In fact, for large probe laser
detuning atom cooling does not offer any advantages in QND
measurements. On the other hand, the presence of even a few torr of
N$_{2}$ gas, which is necessary for optical pumping of
optically-dense vapor, dramatically increases the Lorentzian
linewidth while having little effect on the overall absorption
profile. For example, 10 torr of N$_{2}$, which is the minimum
typically required for quenching \cite{Gay2}, increases the
Lorentzian half-width in Rb by a
factor of 39 from its value in vacuum (3 MHz). In the absence of N$_{2}$,  the effective $%
OD_{L}$ for QND measurements is on the order of $5\times 10^{4}$ for
a 5 cm long cell with atom density of $10^{13}$~cm$^{-3}$. Since one
wins only relatively slowly with $OD$ in quantum measurements, such
large optical densities are crucial to realizing a significant
increase in sensitivity from quantum entanglement.

Different considerations apply for the use of hybrid optical pumping
in spin-exchange relaxation free (SERF) magnetometers \cite{SERF}.
When the magnetometer is operated in steady-state regime the
response to a magnetic field $B$ is given by $S=\gamma BR/(R+\Gamma
)^{2}$. Hence the largest signal is obtained when $R=\Gamma ,$
corresponding to atomic spin polarization of 50\%. Since the atoms
are not fully polarized, the circularly polarized pump laser is
significantly absorbed. This limits the maximum density of the
alkali metal so that the optical depth in the direction of the pump
laser is $OD_{p}\sim 2$.  Using hybrid optical pumping on a
different alkali-metal with much smaller density eliminates this
problem. For light alkali-metals, such as K and Rb, the spin
destruction cross-sections are several orders of magnitude smaller
than the spin-exchange cross-section. Therefore, spin polarization
can be maintained with minor alkali-metal fractions of
$f=10^{-2}-10^{-3}$, increasing operating alkali-metal density or
the size of the cell by a large factor. We have explored this
technique in a $^{21}$Ne-Rb co-magnetometer \cite{Ghosh},  using a
Rb-K mixture with $f_{K}=0.005$ and optical pumping on K. We  obtained an order of magnitude higher polarization of $%
^{21}$Ne than was possible using a single alkali-metal
\cite{Lawrence}, opening new potential for sensitive nuclear spin
gyroscopes \cite{Kornack}.

In conclusion, we describe a simple technique for spin polarization
of optically-thick alkali-metal atoms using spin-exchange with an
optically-thin species. This approach, in combination with high
quality anti-relaxation surface coatings, opens the possibility of
creating very optically-dense spin-polarized alkali-metal vapors
without any quenching gas. Such vapors can be used to transfer
polarization to other species, such as atomic hydrogen or electrons.
They are also useful for spin quantum non-demolition measurements.
Even in the presence of buffer gas, hybrid optical pumping allows
one to independently control absorption of optical pumping light,
which is beneficial in ultra-sensitive atomic magnetometers and
other optical pumping experiments.

We'd like to thank Scott Seltzer for fabrication of the alkali-metal
cells and Ivanna Dimitrova for discussions. This work was supported
by an ONR MURI award.

\end{document}